\documentclass[twocolumn,superscriptaddress,showpacs,pra]{revtex4}
\usepackage{mathrsfs}
\usepackage{amssymb, amsbsy, amsmath, latexsym, dsfont, array, layout, graphicx,mathrsfs,color}

\newcommand{\one}{\mathds{1}}
\newcommand{\ket}[1]{\left|{#1}\right\rangle}
\newcommand{\bra}[1]{\left\langle{#1}\right|}

\begin{document}

\title{Localized State in a Two-Dimensional Quantum Walk on a Disordered Lattice}
\author{Peng Xue\footnote{gnep.eux@gmail.com}
}
\affiliation{Department of Physics, Southeast University, Nanjing
211189, China}
\author{Rong Zhang}
\affiliation{Department of Physics, Southeast University, Nanjing
211189, China}
\author{Zhihao Bian}
\affiliation{Department of Physics, Southeast University, Nanjing
211189, China}
\author{Xiang Zhan}
\affiliation{Department of Physics, Southeast University, Nanjing
211189, China}
\author{Hao Qin}
\affiliation{Department of Physics, Southeast University, Nanjing
211189, China}
\author{Barry C. Sanders}
\affiliation{
    Hefei National Laboratory for Physical Sciences at Microscale and Department of Modern Physics,
    University of Science and Technology of China, Hefei, Anhui 230026, China
    }
\affiliation{
    Shanghai Branch,
    CAS Center for Excellence and Synergetic Innovation Center
        in Quantum Information and Quantum Physics,
    University of Science and Technology of China, Shanghai 201315, China
    }
\affiliation{
    Institute for Quantum Science and Technology, University of Calgary, Alberta, Canada T2N 1N4
    }
\affiliation{
    Program in Quantum Information Science,
    Canadian Institute for Advanced Research,
    Toronto, Ontario M5G 1Z8, Canada
    }
\begin{abstract}
We realize a pair of simultaneous ten-step one-dimensional quantum walks with two walkers sharing coins,
which we prove is analogous to the ten-step two-dimensional quantum walk with a single walker
holding a four-dimensional coin. Our experiment demonstrates a ten-step quantum walk over
an $11\times 11$ two-dimensional lattice with a line defect, thereby realizing a localized walker state.
\end{abstract}


\maketitle

\section{Introduction}
\label{sec:introduction}
Quantum walks (QWs)~\cite{ADZ93},
which are the quantum analogue
of classical random walks (RWs),
are valuable in diverse areas
including quantum
algorithms~\cite{Amb03,CCD+03,SKW03,Kem03a}, quantum
computing~\cite{Chi09,CGW13,LCE+10}, state transfer and
quantum routing~\cite{ZQB+14},
quantum simulation~\cite{SGR+12},
topological phase transiti~\cite{KRBD10,Asb12,KBF+12}, energy transport in
photosynthesis~\cite{OPD06,HSW10}, Anderson
localization~\cite{WLK+12,ZXT14,Seg13,YKE08,Kon10,SK10,COR+13,SCP+11,XQT14,ZX14} and
quantum chaos~\cite{WLK+04,BB04,BNP+06,GAS+13,XQTS14}.
The one-dimensional (1D) QW has been realized with
nuclear magnetic resonance~\cite{DLX+03},
atoms~\cite{ZKG+10,SMS+09,KFC+09,CREG06,XS09,XS08},
and photons~\cite{BMK+99,DSB+05,PLP+08,X15,B15}.
Notably the 1D QW has a classical-wave description~\cite{SCP+10,SSV+12,XS13}
whereas the two-dimensional (2D) QW~\cite{MBSS02,RS05,AAA10}
introduces purely quantum effects~\cite{XS12}.
Consequently,
the 2D QW over integer time~$t$ is of paramount interest motivating
recent photonic realizations~\cite{SGR+12,PLM+10,PKF+14,JDL+13,COR+13}
that are actually constructed with a pair of 1D QWs
and presume a relation between two 1D QWs and one 2D QW.

Here we demonstrate experimentally a QW localized state by realizing a line defect
in the reduced QW position distribution
$\tilde{P}_t^{xy}$ over an $11 \times 11$ 2D $(x,y)$ lattice and compare to the theoretical prediction~$P_t^{xy}$. We use $\tilde{}$ to denote experimental quantities, superscripts ${}^x$ and~${}^y$ to denote lattice sites $x$ and $y$, and subscript~${}_t$ to denote the time index.
The localized state of the walker as a signature of 2D QW localization presents the property as the probability distribution of the walker state is highly localized in certain positions instead of spreading. In additional we prove an isomorphism between a pair of 1D QWs sharing coins~\cite{XS12} and a single 2D QW on an integer-valued Cartesian $(x,y)$ lattice (seeing in Appendix). Our proof of the isomorphism between two walkers in one dimension sharing coins
and one walker in two dimensions with a higher-dimensional coin
makes rigorous an oft-used but previously unproven assumption of this isomorphism.

We evaluate the quality of experimental simulation
in terms of the
time-dependent discrepancy
\begin{equation}
	s_t=\frac{1}{2}\sum_{x,y}\left|\tilde{P}_t^{xy}-P_t^{xy}\right|,
\end{equation}
using the 1-norm distance~\cite{BFL+10} between theoretical
and experimental reduced walker distribution on the 2D lattice.
In particular we show
that the discrepancy~$s_t$ is small for our realization,
indicating a successful experimental simulation of a localized state in a 2D QW.

\section{Theory: Localization in a quantum walk}
\label{sec:theory}

Compared to ballistic QWs,
a walk in a disordered lattice leads to an absence of diffusion,
and the wave function of the walker
becomes localized~\cite{And58}. That is, the walker will be observed in a certain position with high probability instead of spreading ballistically. Thus the localized state of the walker is a good evidence for observing a localized QW.

\begin{figure}
\includegraphics[width=3.5cm]{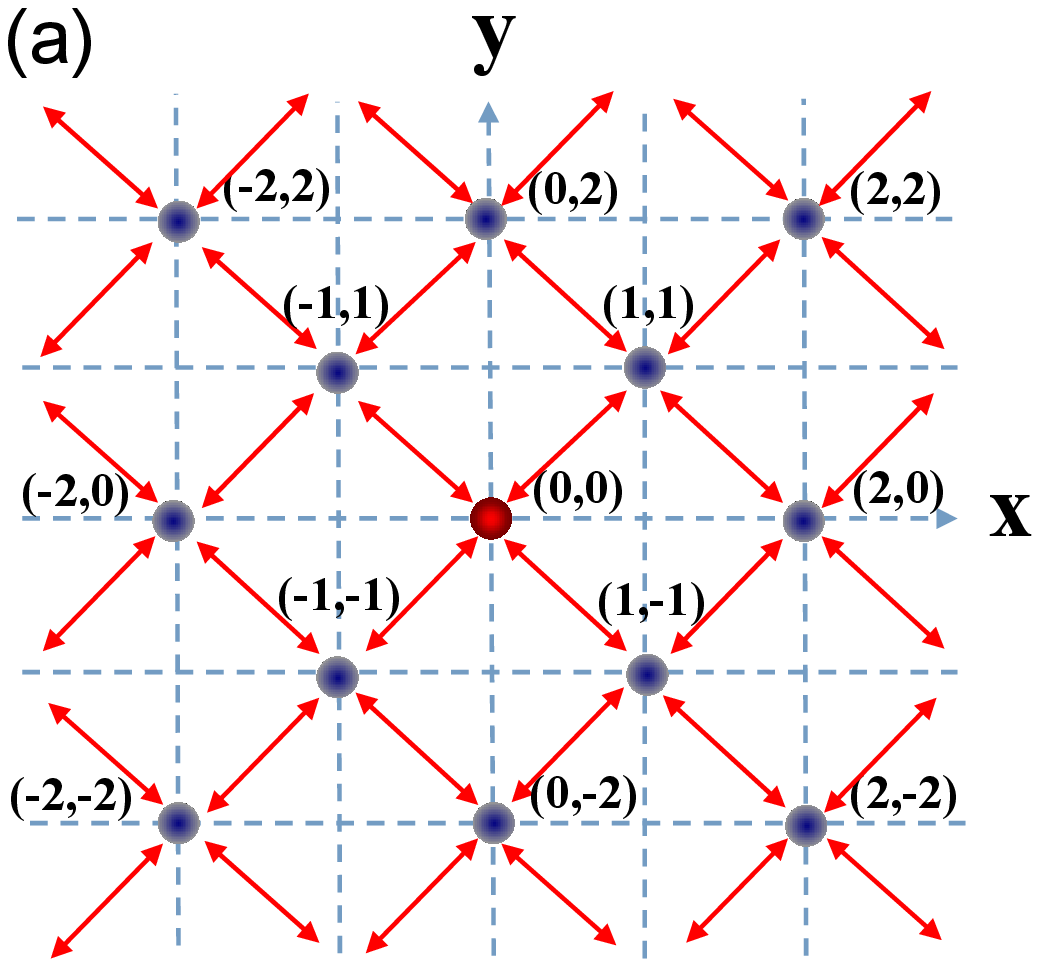}
\includegraphics[width=4.2cm]{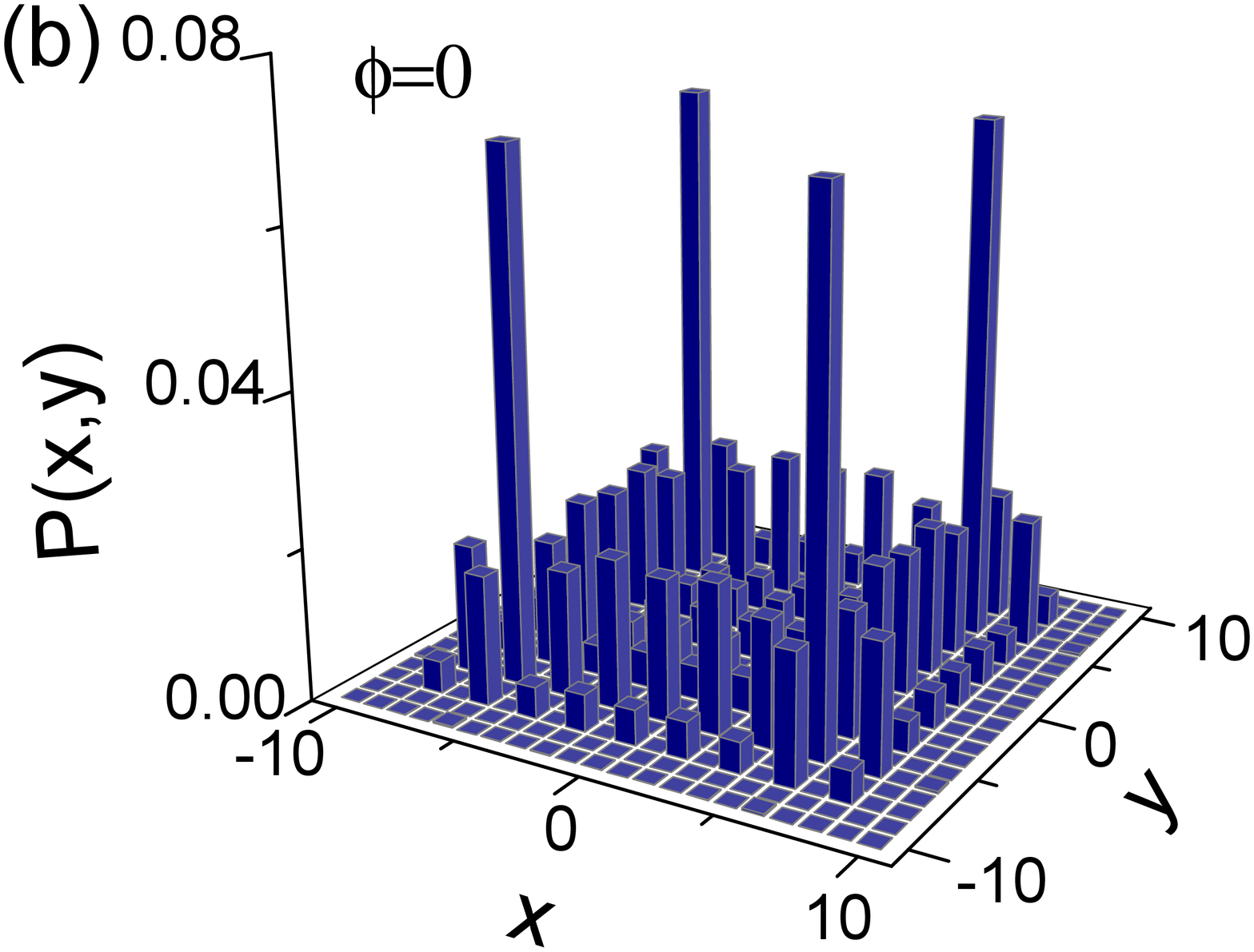}
\includegraphics[width=4.2cm]{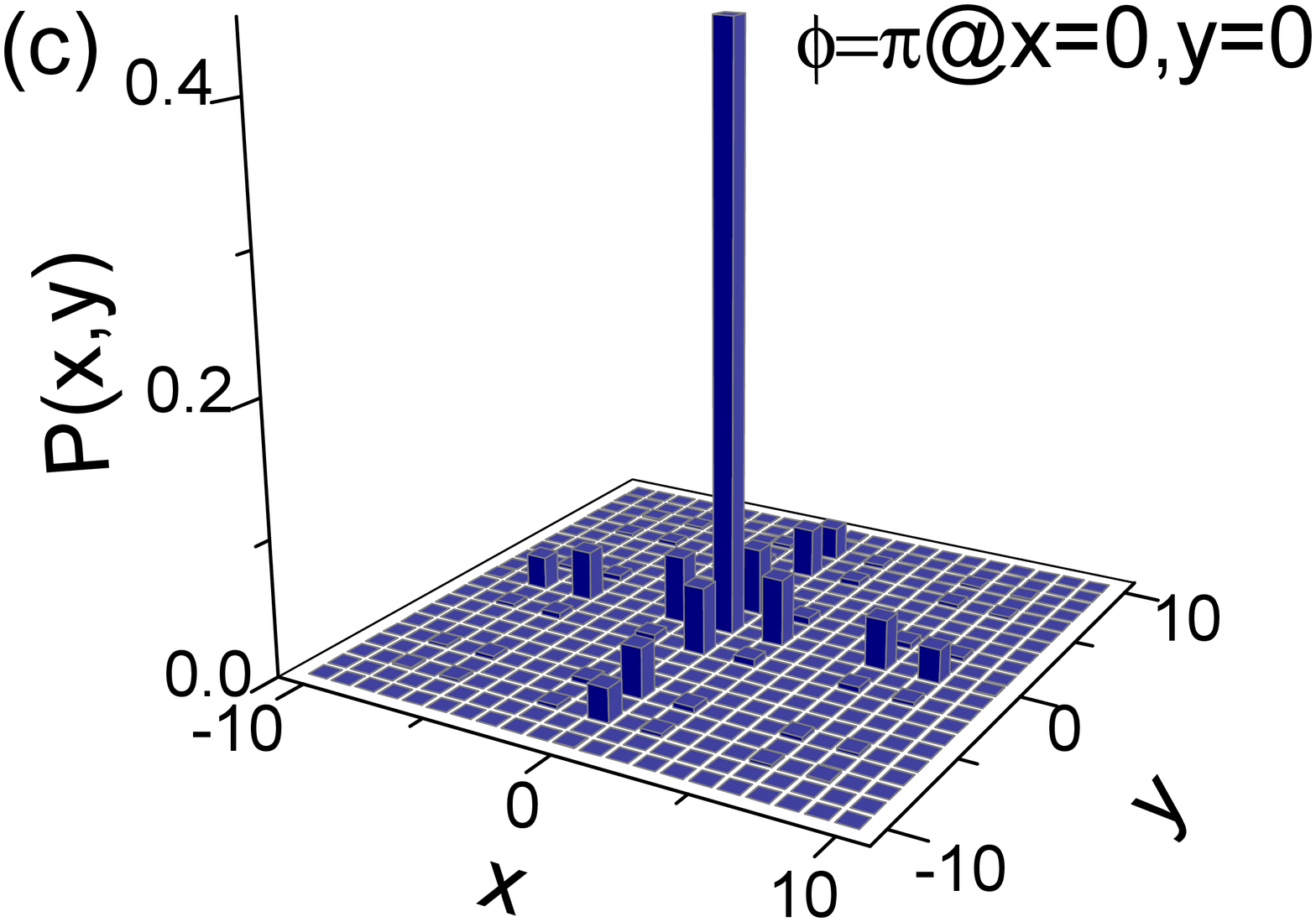}
\includegraphics[width=4.2cm]{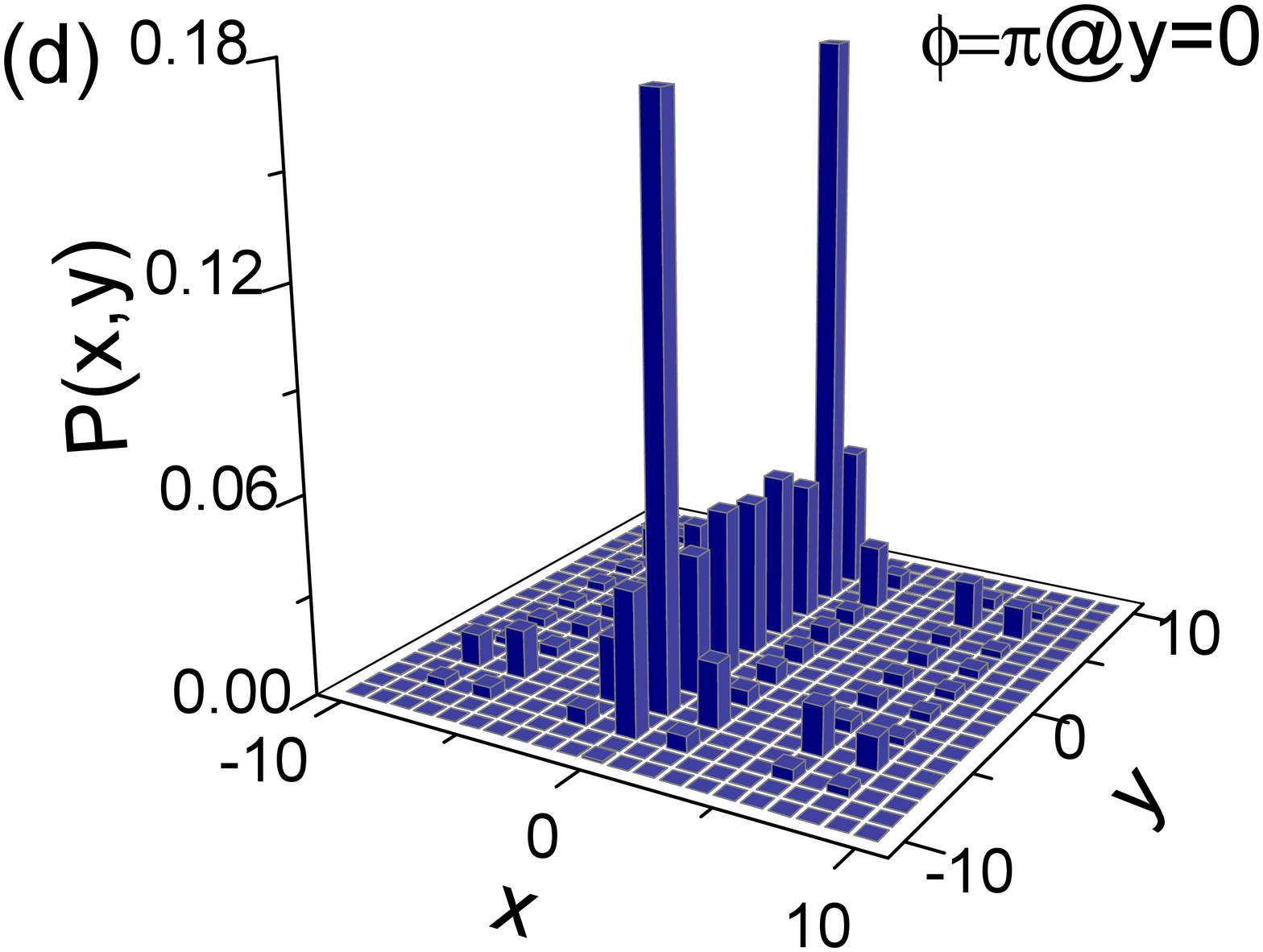}
\caption{%
	(Color online.)
(a)~The 2D lattice of vertices that represent
the state space of two walkers populating
an $11\times 11$ position lattice in an interferometer network.
(b)~Theoretical position distribution after~$10$ steps of a homogenous 2D Hadamard QW.
(c)~Theoretical position distribution after~$10$ steps of a 2D Hadamard QW with line phase
defects $\phi=\pi$ on both $x=0$ and $y=0$. (d) Theoretical position distribution after~$10$ steps of a 2D Hadamard QW with line phase
defects $\phi=\pi$ only on $y=0$.}\label{theory}
\end{figure}

The unitary operation for a single step of QW
in a disordered lattice shown in Fig.~\ref{theory}(a) is
\begin{align}
\label{eq:finitedimensionalV'}
	V_t^{\text{2D}}(\phi)
		=\sum_{x,y\in\Delta_t}\sum_{c,d\in\mathbb{B}}&
			 \text{e}^{\mathrm{i}\phi(x,y)}\ket{x+(-1)^c,y+(-1)^d}\bra{x,y}\nonumber\\&
\otimes\ket{c,d}\bra{c,d}H^{\otimes 2},
\end{align}
where $H=\begin{pmatrix}1&1\\1&-1\end{pmatrix}/\sqrt{2}$ is a Hadamard coin operator. In this paper we consider two types of disorders that are
represented by position-dependent string phase defects
$\text{e}^{\mathrm{i}\phi(\delta_{x,0}+\delta_{y,0})}$ and $\text{e}^{\mathrm{i}\phi\delta_{y,0}}$
with $\delta_{x(y),0}$ the Kronecker $\delta$.

The first type of disorder
corresponds to the case that the first (second) walker is controlled by a
Hadamard coin, walks along $x$ ($y$) direction and obtains an
additional phase $\phi$ whenever passing through $x=0$ ($y=0$).
In contrast the second case corresponds to the case that the second walker obtains
an additional phase whenever passing through $y=0$.
Both cases
break the translational symmetry of the standard QW without creating defects.

Compared to the standard QW, which can be factorized into two
independent distributions of 1D Hadamard QWs as shown in Fig.~\ref{theory}(b),
the 2D QW with position-dependent string phase defect shows a completely
different position distribution as shown in Figs.~\ref{theory}(c) and (d). A QW with
phase defects on $y=0$ is topologically equivalent to that with a walker
on a 2D regular lattice that is trapped on line $x=0$.
On the other hand,
a QW with phase defects on $x(y)=0$, the QW is topologically equivalent
to that with a walker is localized on lines $x(y)=0$. The maximal
probability of the walker appears at the junction point $(0,0)$.

\section{Experiment}
\label{sec:experiment}
Here we simulate experimentally a 2D photonic walk with 1D QW
by realizing two walkers passing through a disordered lattice and employing
the separable coin operation $H^{\otimes 2}$.
We simulate two kinds of disordered lattices:
(i)~a single-point phase defect in the original position $(0,0)$
and (ii)~a string phase defect in the axis $y=0$.
In this way we can observe localization both (i)~on a single point and (ii)~on a line.

\subsection{Positions of one-dimensional walkers}

QWs can be produced by photons passing through a cascade of
birefringent calcite beam displacers (BDs) arranged in a network of
Mach-Zehnder interferometers~\cite{BNP+06,XQTS14,XQT14}. The direction of the single-photon transmission is controlled by the coin
state, i.e., physically the photon polarization.

Each interferometric output corresponds to a given point in the
space and time location of the 1D QW.
Here for 2D QW, pairs of photons are created
via spontaneous parametric down conversion (SPDC) and then injected
separately into the interferometer network from different input
ports. They do not interfere with each other. Pairs of photons
propagate along $x$ and $y$ axes respectively which correspond to
the four different directions taken by single-photon in one step on
a 2D lattice.

In this scenario, disorder can be added in the
evolution by simply inserting polarization-independent phase
shifters (PSs) between the different interferometer arms. Benefiting from the
novel technology of PSs applied in arbitrary positions and stability
of the BD interferometer network, we are able to realize a $10$-step 2D QW within an $11\times 11$ lattice influenced by various types of
controllable disorders.
With this instrument, we observe that photon wave functions
are trapped not only at single points but also on lines.
Furthermore these defects can be used to implement arbitrary phase maps in QWs.

\subsection{One quantum-walk step}

In our experiment the setup in Fig.~\ref{fig:setup} has been realized by using the
BD array as interferometer network similar to that used
in~\cite{XQT14,BNP+06,XQTS14}. By taking advantage of the intrinsically
stable interferometers, our approach is robust and able to control
both coins and walkers at each step. Benefiting from the fully
controllable implementation, we experimentally study the impact of
the position-dependent phase defects on the localization effect in a
QW architecture and the experimental results agree with the
theoretical predictions. Compared to the previous experimental
results which only simulated localization effect by trapping the
walker in a certain single point~\cite{SCP+11,COR+13,XQT14}, we
experimentally localize the walker on the lines instead.

These data are compared to the
theoretical predictions. If the data are not satisfactory with
respect to the 1-norm distance~$s_t$ of the walker distribution,
we discard the data, adjust the PSs and repeat. This
postselection-like method provides an excellent agreement between
the measured probability distribution (measured position variance)
and theoretical prediction.

\subsection{Source and detection}

The photon pairs generated via type-I SPDC in $0.5$mm-thick nonlinear-$\beta$-barium-borate
(BBO) crystal cut at $29.41^o$, pumped by a $400.8$nm CW diode
laser with up to $100$mW of power. For 2D QWs, photon pairs at
wavelength $801.6$nm are prepared into a symmetric initial state
$\left[(\ket{H}+\mathrm{i}\ket{V})/\sqrt{2}\right]^{\otimes2}$ via a
polarizing beam splitter (PBS) following by waveplates. Interference
filters determine the photon bandwidth $3$nm and then pairs of
downconverted photons are steered into the different optical modes
(up and down) of the linear-optical network formed by a series of
BDs, half-wave plates (HWPs) and PSs.

Output photons are detected via avalanche
photo-diodes (APDs) with dark count rate of $<100$s$^{-1}$ whose
coincident signals---monitored using a commercially available
counting logic---are used to postselect two single-photon events.
The total coincident counts are about $300$s$^{-1}$ (the coincident
counts are collected over $60$s). The probability of creating more
than one photon pair is less than $10^{-4}$ and can be neglected.

\begin{figure}
\includegraphics[width=8.5cm]{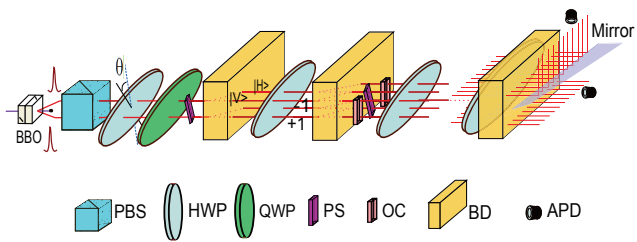}
\caption{(Color online.) Detailed sketch of the setup for $10$-step
2D QW with position-dependent phase defect $\phi$ on $x(y)=0$.
Photon pairs created via type-I SPDC are injected to the optical
network from different ports. Arbitrary initial coin states are
prepared via a PBS, HWP and QWP. PSs are placed in the corresponding
spatial modes and the optical compensators (OCs) are used to
compensate the temporal delay caused by PSs. Coincident detection of
photons at the APDs ($7$ns time window) predicts a successful run of the
QW.}
\label{fig:setup}
\end{figure}

The coin state is encoded in the polarization $\ket{H}$ and
$\ket{V}$ of the input photon. In the basis $\{\ket{H},\ket{V}\}$,
the Hadamard operator is realized with a HWP set to $\pi/8$. The
walkers' positions are represented by longitudinal spatial modes.
The unitary operator shown in Eq.~(1) manipulates the wavepacket to
propagate according to the polarization of the photons. The specific phase $\phi$ can be realized by adjusting the relative angle between the PS
and the following BD.

The spatial modes are separated by a BD with length $28.165$mm and
clear aperture $33$mm$\times 15$mm. After passing a BD, the vertically polarized light is directly transmitted. Whereas the horizontal light undergoes a $3$mm lateral displacement into a neighboring mode. Each pair of BDs forms an interferometer. Only odd (even)
sites of the walker are labeled at each odd (even) step, as the
probabilities of the walker appearing on the other sites are zero.
Pairs of photons are injected from different ports and propagate in
different layers of the BD interferometer network.

The first $10$ steps of the QW with position-dependent phase defect
$\phi$ applied on the two axes $x=0$ and $y=0$ are realized via cascaded interferometric network shown in Fig.~\ref{fig:setup}.
The interference visibility is reached $0.998$ per step. 
The probabilities $P(x,y)$ are
obtained by normalizing photon counts via a coincidence measurement
for two walkers at position $x$ and $y$ for the respective step.

\begin{figure}
\includegraphics[width=8.5cm]{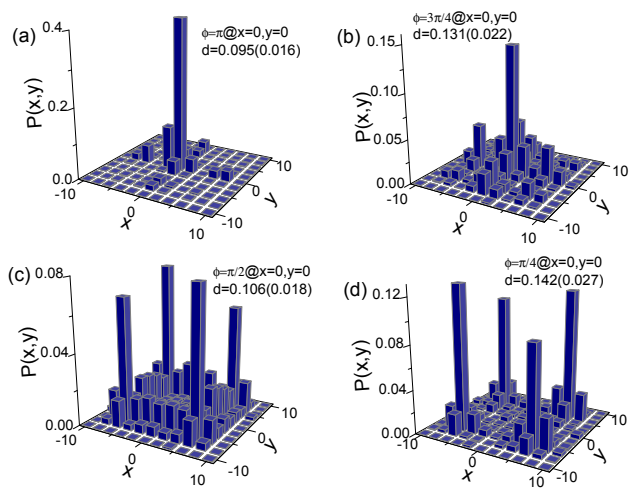}
\caption{Experimental data of probability distributions of the
$10$-step 2D Hadamard QW with position-dependent string phase
defects on both $x=0$ and $y=0$: (a) $\phi=\pi$, (b) $\phi=3\pi/4$,
(c) $\phi=\pi/2$, and (d) $\phi=\pi/4$. The walkers start from the
original position $(0,0)$ with the symmetric coin
state.}\label{pointdefect}
\end{figure}

The measured probability
distributions for $1$ to $10$ steps of a 2D Hadamard QW with
position-dependent phase defect $\phi=\pi$ on $x(y)=0$ and the
symmetric initial coin state are shown in Fig.~\ref{pointdefect}(a).
The 1-norm
distance $0.095\pm0.016$ (after $10$
steps) promises a good agreement between
the experimental results of probabilities and theoretic predictions. The walkers' state after $10$ steps clearly shows the
characteristic shape of a localization distribution: the wave
functions of photons are trapped on two axes $x=0$ and $y=0$, and a
pronounced peak of the probability $0.424\pm0.015$ (with theoretical
prediction $0.441$) in the junction point $(0,0)$, which displays the signature of the localization effect. In contrast to
the ideal standard 2D Hadamard QW, the expansion of the wavepacket is
highly suppressed and the probabilities $P(x,0)$ and $P(0,y)$ are
enhanced.

\subsection{Results}

Our experimental result highlights the full control of the
implementation of the 2D QW. In Fig.~\ref{pointdefect}, we show the impact of phase
defects $\phi\in\left[0,\pi\right]$ on the localization effect. Figs.~\ref{pointdefect}(b) and (d) show the position
distribution of the $10$-step 2D Hadamard QW with $\phi=3\pi/4,
\pi/2, \pi/4$. For the symmetric initial coin state, the two walkers
behave same and show the symmetric distributions.

The localization
effect can be observed in the range $\phi\in\left[3\pi/4,\pi\right]$,
and the recurrence probability $P_{10}(0,0)$ increases with $\phi$,
which agrees with the analytic result. Especially for $\phi=\pi$ the
walkers are almost completely trapped on the axes $x$ and $y$. If
$\phi$ decreases, the 2D QW's behaviour tends to be ballistic. For
$\phi=\pi/2$ the wave functions of photons are distributed as same
as standard Hadamard QW without phase defects. For $\phi=\pi/4$, the
photons spread even faster and show a super-ballistic behaviour. Thus,
whether or not the localization effect can be observed depends
on the choices of phase defects~\cite{WLK+12,ZXT14}.

\begin{figure}
\includegraphics[width=8.5cm]{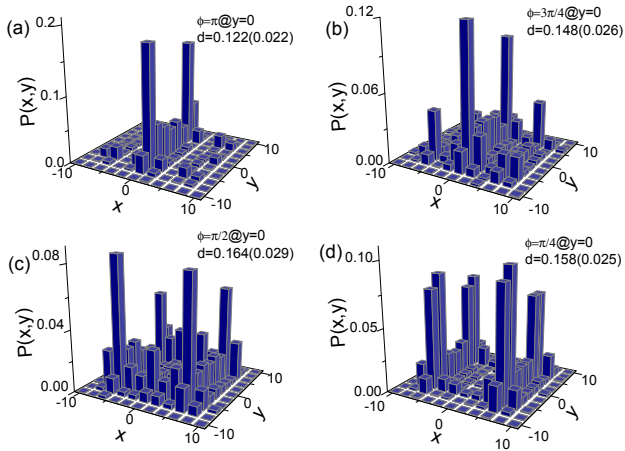}
\caption{Experimental data of probability distributions of the
$10$-step 2D Hadamard QW with position-dependent string phase
defects only on $y=0$: (a) $\phi=\pi$, (b) $\phi=3\pi/4$, (c)
$\phi=\pi/2$, and (d) $\phi=\pi/4$.}\label{linedefect}
\end{figure}

Now we add the phase defects only on $y$ axis.
That is, if and only
if the walker who walks along the $y$ axis arrives at $y=0$
obtains an additional phase $\phi$. Experimentally we rearrange the PSs and
photons propagating in the lower layer pass through the PSs. In
Fig.~\ref{linedefect} we show the measured position distribution of 2D QW with the
string phase defects,
which displays that the photons appear on a
line with relative large probabilities.

\begin{figure}
\includegraphics[width=4.2cm]{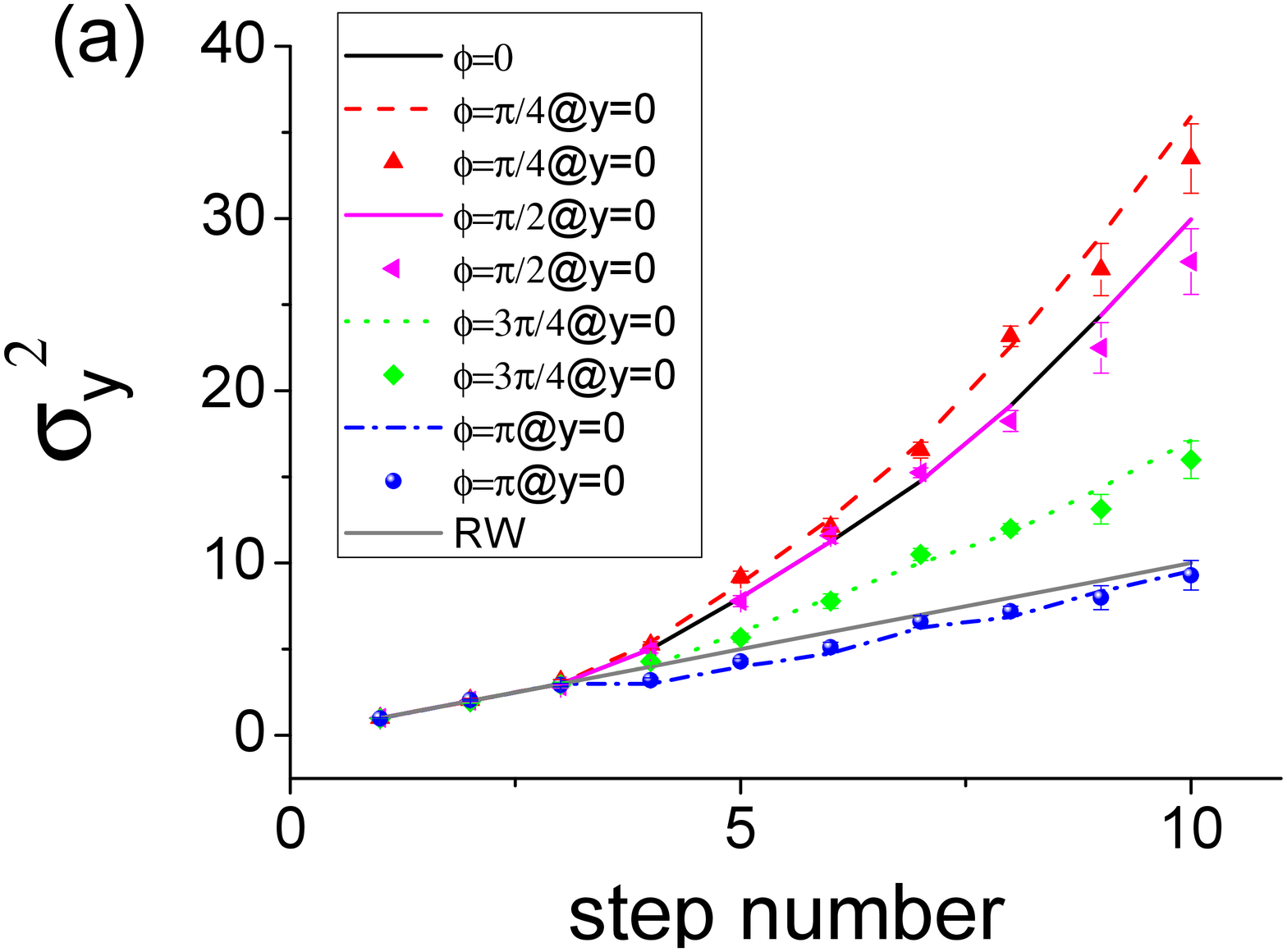}
\includegraphics[width=4.2cm]{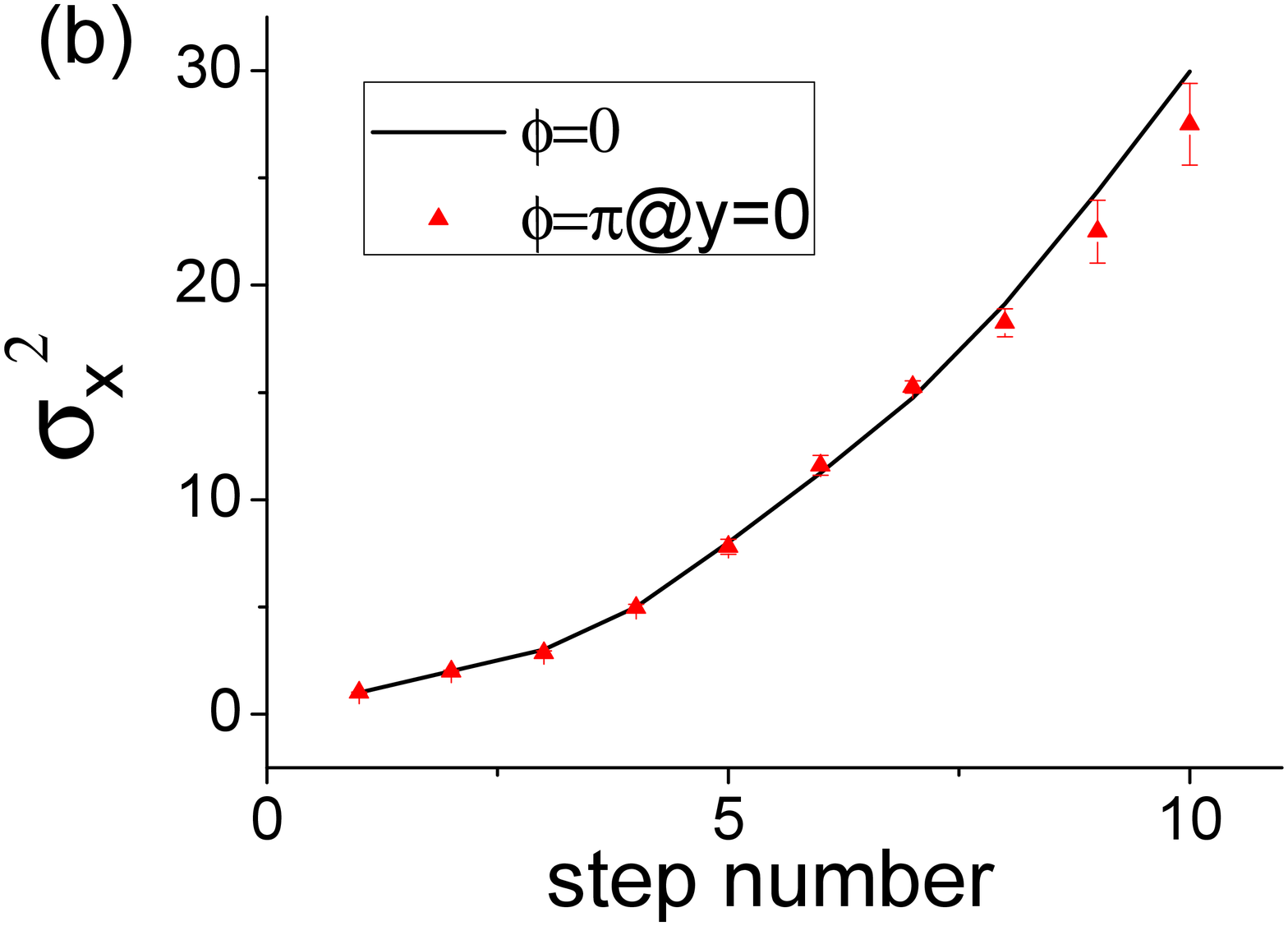}
\caption{(Color online.) (a)~Measured trend of the variance of the
walker who walks along $y$ axis up to $10$ steps with respective
theoretical predictions (lines). (b)~Measured dynamics evolution of
the position variance of the walker who walks along $x$ axis. As
the phase defects are only applied on $y=0$, the walker along $x$
axis is not affected. Thus for all $\phi$ the walker shows a
ballistic behaviour.}\label{variance}
\end{figure}

Thus, the photons are
localized on the $x$ axis for $\phi$ large enough.
On the $x$ axis, the
photon distribution is similar to that of the 1D standard Hadamard QW. In
Fig.~\ref{variance}, measured position variances of the walker along $y$ axis
show the impact of phase defects. For $\phi=\pi/2$
photons show a ballistic behaviour. For $\phi=\pi/4$ they move even
faster and show a super-ballistic behaviour. For $\phi=3\pi/4$
and $\phi=\pi$ they stagnate and show localization. For $\phi=\pi$ the variances are even smaller than those of
the classical RW. Whereas the walker walking along $x$ axis is not
influenced.

\section{Conclusions}
\label{sec:conclusions}

Our experimental architecture can be generalized to more than two
dimensions with the same BD interferometer network, a
deterministic multi-photon source and joined multi-photon
measurement. Multiple photons undergoing an interferometer network
represent the walker in higher-dimensional structures and the
polarization of the photons represent the coins manipulated by the
waveplates. This opens a large unexplored field of research such as
quantum simulation with multiple walkers.

In summary, we implement a stable and efficient way to realize 2D QW
embedded in a broader framework and show the position-dependent
phase defects can influence the evolution of wavepackets.
The 2D QW
with string phase defects has the wave functions of photons
localized in the certain lines. This is for the first time we
observe localization on the lines instead of single points.

\acknowledgements
This work has been supported by NSFC under 11174052 and 11474049, and CAST Innovation fund.
BCS appreciates financial support from the 1000 Talent Plan of China,
AITF and NSERC.

\begin{appendix}
\section{Isomorphism between two one-dimensional quantum walks and one two-dimensional quantum walk}
In this section we begin by describing the 1D QW,
then describe a pair of 1D QWs with a shared coin~\cite{XS12}
and follow with a discussion of the 2D QWs.
Following these descriptions,
we prove an isomorphism between a pair of 1D QWs sharing a quantum coin~\cite{XS12}
and the 2D QW.

\subsection{One-dimensional quantum walk}
The 1D QW has a walker moving along an integer lattice whose sites are indexed by
$x\in\mathbb{Z}$.
Thus, the basis set for the walker state is $\{\ket{x};x\in\mathbb{Z}\}$.
The coin operator~$C^\text{1D}$ is an element of the Lie Group $SU(2)$ and can be site-dependent,
which is important for introducing lattice defects.
Therefore, we write the coin operator as
\begin{equation}
\label{eq:C1D}
	C^\text{1D}:=\sum_{x\in\mathbb{Z}}\ket{x}\bra{x}\otimes C^x
\end{equation}
to present site-dependent coin operation which is used widely in realizing generalized measurement via QW~\cite{B15}. Whereas non-site-dependent coin operation can be written as $\one\otimes C^x$ with $C^x\in SU(2)$ uniform for arbitrary $x$.

The coin-state basis is
\begin{equation}
	\{\ket{c}\in P\mathbb{C}^2;c\in\mathbb{B}\}
\end{equation}
for~$\mathbb{B}=\{0,1\}$ the bit space and $P\mathbb{C}^2$ the projective space of pairs of complex numbers.
Thus, we can write
\begin{equation}
\label{eq:Cx}
	C^x=\begin{pmatrix}
			 \text{e}^{-\text{i}\varphi^x}\cos\theta^x&\text{e}^{\text{i}\psi^x}\sin\theta^x\\
			 \text{e}^{-\text{i}\psi^x}\sin\theta^x&\text{e}^{\text{i}\varphi^x}\cos\theta^x
		\end{pmatrix}
\end{equation}
being the $2\times 2$ complex matrix representation for $SU(2)$,
which is parameterized by three independent
$x$-dependent angles $\theta^x$, $\psi^x$ and $\varphi^x$.

The QW step operator~$U$ is obtained by combining the coin flip
with the conditional translation of the walker.
The conditional translation operator is
\begin{equation}
\label{eq:1Dconditionaltranslation}
	 T^{\text{1D}}=\sum_{x\in\mathbb{Z}}\left(\ket{x}\bra{x+1}\otimes\ket{0}\bra{0}
		+\ket{x+1}\bra{x}\otimes\ket{1}\bra{1}\right).
\end{equation}
The unitary QW step operator is thus
\begin{equation}
\label{eq:U=TC}
	U^{1\text D}=T^{\text{1D}}C^\text{1D}.
\end{equation}
The walker's evolution is obtained in discrete steps with evolution time given
by
\begin{equation}
\label{eq:1Dtbound}
	t\in\mathbb{N}=\{0,1,2,\dots\},
\end{equation}
and the evolution at time~$t$ is given by~$(U^{1\text D})^t$.

For fixed~$t$, and for a walker whose state has support over a finite domain of $\{x\in\mathbb{Z}\}$,
the step operator~$U^{1\text D}$ has a finite-dimensional representation.
For the initial walker state commencing as a wholly localized state at the origin $x=0$,
the domain at time~$t$ can be restricted to
\begin{equation}
\label{eq:Delta1D}
	x\in\Delta_t:=\{-t,\dots,t\}.
\end{equation}
(Actually the domain can be restricted to even and odd sublattices depending on the parity of~$t$,
but we ignore this simplification here.)

The 1D QW unitary step operator~(\ref{eq:U=TC}) can be expressed as
\begin{align}
\label{eq:finitedimensionalU}
	U_t^{\text{1D}}
		 =&\sum_{x\in\Delta_t}\big(\ket{x+1}\bra{x}\otimes\ket{0}\bra{0}
					\nonumber
			+\ket{x-1}\bra{x}\otimes\ket{1}\bra{1}\big)
					\nonumber\\&\times
			\sum_{x'\in\Delta_t}\ket{x'}\bra{x'}
			 \otimes C^{x'}
					\nonumber\\
		 =&\sum_{x\in\Delta_t}\sum_{c\in\mathbb{B}}
		 	\ket{x+(-1)^c}\bra{x}\otimes(\ket{c}\bra{c} C^x)
\end{align}
where we have employed the periodic boundary condition
\begin{equation}
	\ket{\pm(t+1)}\equiv\ket{\mp t}.
\end{equation}
For
\begin{equation}
\label{eq:d1D}
	d_t^{1\text D}:=2(2t+1),
\end{equation}
the operator $U$~(\ref{eq:finitedimensionalU}) can be expressed
as a $\left(d_t^{1\text D}\times d_t^{1\text D}\right)$-dimensional special unitary matrix.

\subsection{Two one-dimensional quantum walks}

Now let us consider two 1D QWs, each holding a coin with site-dependent $SU(2)$ operator.
If the two QWs are completely independent of each other,
the evolution is simply a power~$t$
of the tensor product of individual evolutions: $(U_t^{1\text D}\otimes U_t^{1\text D})^t$,
which can be expressed as a
special unitary matrix of dimension
\begin{equation}
\label{eq:dimension}
	\left(d_t^{1\text D}\right)^2\times \left(d_t^{1\text D}\right)^2.
\end{equation}

The two-walker step-by-step unitary evolution operator is
\begin{align}
\label{eq:finitedimensionalV}
	U_t^\text{1D1D}&=T^\text{1D1D} C^\text{1D1D}\nonumber\\
		&=\sum_{x,y\in\Delta_t}\sum_{c,d,\in\mathbb{B}}
			\ket{x+(-1)^c,y+(-1)^d}\bra{x,y}\nonumber\\
				&\otimes(\ket{c,d}\bra{c,d}C^{xy}),
\end{align}
where
\begin{align}
T^\text{1D1D}=&\sum_{x,y\in\mathbb{Z}}\sum_{c,d,\in\mathbb{B}}
			\ket{x+(-1)^c,y+(-1)^d}\bra{x,y}\nonumber\\
&\otimes\ket{c,d}\bra{c,d},
 \end{align}
 \begin{equation}C^\text{1D1D}=\sum_{x,y\in\mathbb{Z}}\ket{x,y}\bra{x,y}\otimes C^{xy}\end{equation}
and
\begin{equation}
\label{eq:C4}
	C^{xy}\in SU(4).
\end{equation}
This coin operator can be parameterized by fifteen independent angles,
and this operator~(\ref{eq:finitedimensionalV}) reduces to
$U_t^{\text{1D}}\otimes U_t^{\text{1D}}$ if
\begin{equation}
\label{eq:C2x2}
	C^{xy}=C^x\otimes C^y\in SU(2)\times SU(2).
\end{equation}
Two independent walkers thus necessarily remain independent under this factorizable evolution.

If the coin operator~(\ref{eq:C4}) is not factorizable,
two walkers can become entangled by sharing coins,
which is achieved by a fractional-swap operation
\begin{align}
\label{eq:CXi}
	C^{xy}
		=&\Xi^{\tau^{xy}}
					\nonumber\\
		=&\frac{1}{2}
			\begin{pmatrix}
				2&0&0&0\\0&1+(-1)^{\tau^{xy}}&1-(-1)^{\tau^{xy}}&0\\
				0&1-(-1)^{\tau^{xy}}&1+(-1)^{\tau^{xy}}&0\\0&0&0&2
			\end{pmatrix}
\end{align}
for $\Xi$ the swap operator and $\tau^{xy}\in(0,1)$~\cite{XS12}.
If the walkers' coin-sharing procedure is independent of position,
then $\tau^{xy}\equiv \tau$ (a constant).
Thus, $U_t^\text{1D1D}$~(\ref{eq:finitedimensionalV})
can be expressed as a
special unitary matrix of dimension $\left(d_t^{1\text D}\right)^2\times \left(d_t^{1\text D}\right)^2$ as same as~(\ref{eq:dimension}).

\subsection{Two-dimensional quantum walk}

For a a single quantum walker moving along a 2D Cartesian lattice,
a convenient basis choice is
\begin{equation}
	\left\{\ket{x,y,c};(x,y)
		\in\mathbb{Z}^2,c\in\mathbb{B}^2\right\}.
\end{equation}
Thus, the walk is over the 2D integer lattice and the coin-state parameter is given by a two-bit string.

Analogous to the 1D coin operator~(\ref{eq:C1D}),
the 2D coin operator is
\begin{equation}
\label{eq:C2D}
	C^\text{2D}:=\sum_{(x,y)\in\mathbb{Z}^2}\ket{x,y}\bra{x,y}\otimes C^{xy}
\end{equation}
to present 2D site-dependent coin operator. Whereas the non-site-dependent coin operation can be written as $\one\otimes C^{xy}$ with $C^{xy}$ uniform for any $(x,y)$.
Following the coin flip,
translation takes place,
which is given by the 2D translation operator
\begin{align}
\label{eq:2Dconditionaltranslation}
	 T^{\text{2D}}=&\sum_{(x,y)\in\mathbb{Z}^2}\big(\ket{x,y}\bra{x+1,y}\otimes\ket{0,0}\bra{0,0}
				\nonumber\\&
		+\ket{x,y}\bra{x,y+1}\otimes\ket{0,1}\bra{0,1}
				\nonumber\\&
		+\ket{x,y+1}\bra{x,y}\otimes\ket{1,0}\bra{1,0}
				\nonumber\\&
		+\ket{x+1,y}\bra{x,y}\otimes\ket{1,1}\bra{1,1}\big).
\end{align}
The unitary QW step operator is thus $U^{2\text D}=T^{\text{2D}}C^\text{2D}$
analogous to the 1D translation operator~(\ref{eq:1Dconditionaltranslation})
and can be expressed as a $(d_t^{2\text D}\times d_t^{2\text D})$-dimensional special unitary matrix for
\begin{equation}
\label{eq:d2D}
	d_t^{2\text D}:=\left[2(2t+1)\right]^2=\left(d_t^{1\text D}\right)^2.
\end{equation}
The quantum walker accesses only the sub lattice~$\Delta_t^{\otimes 2}$,
which is a two-fold tensor product of the 1D sub lattice~(\ref{eq:Delta1D}).

\subsection{Isomorphism between two one-dimensional quantum walks and one two-dimensional quantum walk}

The isomorphism between two 1D quantum walkers and one 2D quantum walker
is proven if the two transformations are identical in appropriate bases.
We know from Eq.~(\ref{eq:d2D}) that the two matrices have the same size
so the approach in this subsection is to find the appropriate basis transformation
from 1D to 2D so the matrix representations are identical.
Then we need to establish that the transformation~(\ref{eq:xymapping})
and the two-coin operation including fractional swap~(\ref{eq:CXi})
leads to the same unitary step-operator matrix for the two cases of two 1D QWs and one 2D QW.
We show this isomorphism by proving that
\begin{equation}
	U_t^\text{1D1D}=U_t^\text{2D}
\end{equation}
after implementing the co\"{o}rdinate transformation~(\ref{eq:xymapping})
and the fractional quantum-coin swap~(\ref{eq:CXi}).

We choose the mapping
\begin{equation}
\label{eq:xymapping}
	x\mapsto x+y,\;
	y\mapsto x-y,
\end{equation}
to carry co\"{o}rdinates~$x$ and~$y$ for the two 1D walkers
to the joint co\"{o}rdinate of the 2D quantum walker.
Under the transformation~(\ref{eq:xymapping}),
the 2D translation operator~(\ref{eq:2Dconditionaltranslation})
can be rewritten as
\begin{align}
	T^{\text{2D}}
		=\sum_{x,y\in\mathbb{Z}}\sum_{c,d\in \mathbb{B}}
			&\ket{x+(-1)^{c},y+(-1)^d}\bra{x,y}
					\nonumber\\&
			\otimes \ket{c,d}\bra{c,d},
\end{align}
which evidently matches $T^\text{1D1D}$---a crucial part of $U_t^{\text{1D1D}}$ in Eq.~(\ref{eq:finitedimensionalV}).
The next step to proving the isomorphism is to decompose the $SU(4)$ coin operator~(\ref{eq:C2D})
according to~\cite{KC01,KBG01,FRS05}
\begin{align}
	C^{xy}=&(u_1\otimes u_2)\big[(Z\otimes X)\Xi^\gamma(Z\otimes \mathds{1})\nonumber\\&
		\times\Xi^\beta(\mathds{1}\otimes X)\Xi^\alpha\big](v_1\otimes v_2)
\label{eq:Cxy}
\end{align}
with Pauli matrices
\begin{equation}
	X=\begin{pmatrix}0&1\\1&0\end{pmatrix},\;
	Z=\begin{pmatrix}1&0\\0&-1\end{pmatrix},
\end{equation}
general $SU(2)$ elements~$u_{1,2}$ and~$v_{1,2}$,
and $\Xi^i$ ($i=\alpha,\beta,\gamma\in[0,1]$) the fractional-swap operation~(\ref{eq:CXi}).
That is,
an arbitrary $SU(4)$ operation on a four-sided coin can be decomposed into
three $\Xi^i$ gates and single-qubit gates.
An arbitrary $SU(4)$ coin can be either separated or entangled. For the former case,
$C^{xy}$ can be decomposed into single-qubit gates only, i.e.,
\begin{equation}
	C^{xy}=C^x\otimes C^y,
\label{eq:separated}
\end{equation}
if and only if
\begin{equation}
	C^x=u_1v_1,\;
	C^y=u_2v_2, \;\alpha=\beta=\gamma=0.\end{equation}
For the latter case, $C^{xy}$ can be decomposed by three $\Xi^i$ gates, i.e. \begin{equation}
	C^{xy}=\Xi^{\tau^{xy}}
\label{eq:entangled}
\end{equation}
if and only if
\begin{equation}
	\beta=\gamma=-1,\;
	\alpha=\tau^{xy},\;
	u_{1,2}=v_{1,2}=\mathds{1}.
\end{equation}
Thus we show the isomorphism between two 1D QWs with two walkers having separated coins and 2D QW with one walker controlling a four-side coin in Eq.~(\ref{eq:separated}), and the isomorphism between two 1D QWs with two walkers sharing coins and 2D QW with one walker controlling a four-side coin in Eq.~(\ref{eq:entangled}) by proving $U_t^\text{1D1D}=U_t^\text{2D}$ for the two cases respectively.

Therefore, a 2D QW with one walker controlled by a $SU(4)$ coin flipping and a 1D QW with two walkers sharing coins~\cite{XS12} is proven to be isomorphic.
Thus, one can use a 1D QW with two walker
to simulate 2D QW if the two walkers share their coins except for the local rotations.

Here we simulate a 2D walk with two 1D quantum walkers
and treat the simple coin flipping operator~$H^{\otimes 2}$
for Hadamard
operator $H$ which is a special case of the coin operators for two 1D QWs~(\ref{eq:C2x2}).
In this case, the above coin operator for two 1D QWs is equivalent to that for a 2D QW $C^{xy}$ in Eq.~(\ref{eq:Cxy}) once
\begin{equation}
	u_1=u_2=H,\;
	\alpha=\beta=\gamma=0,\;
	v_1=v_2=\mathds{1}
\end{equation}
are satisfied.
\end{appendix}

\end{document}